\documentclass[]{raa}  %linenumbers          % referee version: for submission

\usepackage{graphicx,times}             %for PS/EPS graphics inclusion, new
\usepackage{natbib}
\usepackage{amssymb,amsmath}
\bibpunct{(}{)}{;}{a}{}{,}
\usepackage[pagebackref=true]{hyperref}
\usepackage{threeparttable}
\usepackage{xcolor}

\hypersetup{colorlinks=true,citecolor=blue}

\begin{document}

  \title{Identifying New $\gamma$-Ray Sources in All-Sky Surveys Based on Fermipy's Advanced Algorithm}
%   \subtitle{I. Place Your Subtitle Here}

   \volnopage{Vol.0 (20xx) No.0, 000--000}      %%preserved for Editor. DOn't remove!
   \setcounter{page}{1}          %%starting page, preserved for Editor. DOn't remove!

   \author{Yunchuan Xiang %(周爱英) %% Put your Chinese name in "( )" if you like. Note to open line 11 "\usepackage[UTF8]{ctex}"
      \inst{1,2}
   \and Peng Feng
      \inst{1}
   \and Xiaofei Lan
      \inst{1}
   }
%% Here is an example of three authors come from different institutes.
%% For single author or all the authors from an institute, use "\inst{}" only

   \institute{School of Physics and Astronomy, China West Normal University, Nanchong 637009, People's Republic of China; {\it xiang\_yunchuan@yeah.net}\\
%% Please give the E-mail address of the author, to whom future correspondence and
%% offprint requests will be sent.
        \and
             Department of Astronomy, Yunnan University, and Key Laboratory of Astroparticle Physics of Yunnan Province, Kunming, 650091, China;\\
\vs\no
   {\small Received xx month day; accepted xx month day}}

\abstract{We employ an efﬁcient method for identifying $\gamma$-ray sources across the entire sky, leveraging advanced algorithms
from Fermipy, and cleverly utilizing the Galactic diffuse background emission model to partition the entire sky into
72 regions, thereby greatly enhancing the efﬁciency of discovering new sources throughout the sky through multi-
threaded parallel computing. After conﬁrming the reliability of the new method, we applied it for the ﬁrst time to
analyze data from the Fermi Large Area Telescope (Fermi-LAT) encompassing approximately 15.41 yr of all-sky
surveys. Through this analysis, we successfully identiﬁed 1379 new sources with signiﬁcance levels exceeding 4$\sigma$,
of which 497 sources exhibited higher signiﬁcance levels exceeding 5$\sigma$. Subsequently, we performed a systematic
analysis of the spatial extension, spectra, and light variation characteristics of these newly identiﬁed sources. We
identiﬁed 21 extended sources and 23 sources exhibiting spectral curvature above 10 GeV. Additionally, we
identiﬁed 44 variable sources above 1 GeV. 
}

   \authorrunning{ }            %author_head in even pages
   \titlerunning{ }  % title_head in odd pages

   \maketitle
%% The author head (on even pages) and the title head (on odd pages) will be
%% automatically extracted from \author{} and \title{}. Whenever the title is too long,
%% you will be asked to supply a shorter one by inserting either \authorrunning{} or
%% \titlerunning{} before \maketitle. Anyway, you can specify your own heads.
%%
%%
%% Note: In the following text body of your manuscript, please note several differences from
%%       other major journals:
%% (1) \subsection{Please Capitalize the First Letter of Each Notional Word in Subsection Title}
%% (2) Please Capitalize the First Letter of Each Notional Word in all tables' captions

%
%________________________________________________ sections below
%

\section{Data Availability} \label{sec5}
The data supporting the ﬁndings of this study, including the
analysis results such as 4FGL-Xiang.ﬁts, 4FGL-Xiang\_Region\_File.reg, Table \ref{compar_14_14}, the TS maps of 72 ROIs, and other relevant
supplementary materials, have been uploaded to the China-VO
PaperData service at  \href{https://nadc.china-vo.org/res/r101462/}{https://nadc.china-vo.org/res/r101530/}.  We will promptly update the relevant content based on readers' feedback, with the most recent database link taking precedence over the arXiv version. 
Please read the Readme.txt before using or referring to these data. 

\section{Introduction}           %% first-level sections will be auto-capitalized
\label{sect:intro}
Since its launch in 2008, the Large Area Telescope (LAT) aboard the Fermi $\gamma$-ray Space Telescope has revolutionized our understanding of the $\gamma$-ray sky \citep{Abdo2010}.
Fermi-LAT has provided unprecedented sensitivity and resolution, enabling the identification of thousands of $\gamma$-ray sources to date \citep{Abdollahi2020}. 
Despite the groundbreaking progress made by Fermi-LAT in exploring new sources, the GeV emissions of many potential $\gamma$-ray candidates remain unidentified. Furthermore, as data continues to accumulate, the discovery of new high-energy $\gamma$-ray sources is anticipated. To effectively analyze Fermi-LAT data and uncover these new sources, the adoption of advanced data analysis tools is essential. One such tool is Fermipy, a Python package designed to optimize the analysis of Fermi-LAT data \citep{Wood2017}.
 Fermipy integrates with the Fermi Science Tools and provides a user-friendly interface for performing tasks such as source detection, spectral fitting, variability analysis, and spatial analysis.
In this study, we leverage Fermipy to conduct a comprehensive search for new $\gamma$-ray sources across the entire sky. By employing sophisticated statistical methods and robust background models, we aim to discover more new sources and improve our understanding of the $\gamma$-ray sky.

The phenomenon is common in galaxies, particularly in celestial bodies with large-scale structures producing high-energy radiation. 

 In Fermi-LAT 14-year Source Catalog (4FGL-DR4), galaxies with high-energy radiation include normal galaxies, radio galaxies, starburst galaxies, and star-forming galaxies  \citep{Ballet2024}. 
The high-energy radiation from these galaxies may stem from massive stars depleting their core fuel, leading to supernova explosions that release vast amounts of energy and matter.
The energy and matter interact with the surrounding medium to produce high-energy radiation \citep{Woosley2002,Woosley1995}.

Furthermore, active galactic nuclei (AGNs) are the most common $\gamma$-ray sources outside the Milky Way. There are many types of AGNs, including blazar, BL Lac, quasar, and Seyfert galaxy in 4FGL-DR4. The mechanisms that generate high-energy radiation may originate from internal jetting \citep{Guepin2018}, magnetic field acceleration \citep{Katsoulakos2018}, and collisional acceleration \citep{Schlickeiser1996}.

Various stars, characterized by their small-scale structure, are common $\gamma$-rays sources. The currently detected types include supernova remnants (SNR)  \citep{Acero2016}, pulsars \citep{Smith2023}, pulsar wind nebulae \citep{Grondin2011}, and $\gamma$-ray bursts  \citep{Roberts2018}.
The mechanism for generating their high-energy radiation may come from different forms of particle acceleration and interaction processes. For example, the high-energy radiation of supernova remnants may originate from the diffusion shock acceleration mechanism \citep{Bell1978}, the high-energy radiation of pulsars may come from the acceleration of the internal strong magnetic field \citep{Gunn1969}, the rotation of the internal pulsar may power the high-energy radiation of pulsar wind nebulae, and the interaction between the pulsar wind and interstellar medium \citep{Zhang2008}. 
 The high-energy radiation from short-duration $\gamma$-ray bursts is generally believed to originate from the merger of two neutron stars. Neutron stars are compact remnants of stars with strong gravity. 
When two neutron stars merge, their gravitational interactions trigger violent collisions and produce high-energy radiation \citep{Hartley2017,Goldstein2017}.
The origin of high-energy radiation in long-duration $\gamma$-ray bursts is commonly attributed to the collapse of massive star cores \citep{Woosley2006}.

With the continuous accumulation of Fermi-LAT data and the improvement of data analysis technology, high-energy radiation from exotic celestial bodies and special regions, such as variable stars \citep{delValle2014}, binary stars \citep{Harvey2021,DeMartino2013}, molecular clouds \citep{deBoer2017}, and HII regions \citep{Peron2022,Liu2022}, have attracted widespread attention from astronomers. 
The high-energy radiation of variable stars usually originates from stellar activities, including interactions with stellar winds and interstellar media   \citep{delValle2014}, magnetization \citep{Tchekhovskoy2015}, and interactions with companion stars \citep{Cheung2022}. 
Binary stars consist of two celestial bodies, and when one of the binary stars,  such as a neutron star or black hole, accretes matter from its surroundings, this matter is accelerated to an extremely high velocity, generating intense $\gamma$-ray radiation \citep{Narayan1992}. 
Molecular clouds have large cloud-like structures mainly composed of hydrogen molecules and traces of other molecules. High-energy radiation can be observed in regions with molecular clouds; however, this radiation usually does not originate from the molecular cloud but from the celestial bodies surrounding the molecular cloud. For example, high-energy cosmic rays from surrounding celestial bodies collide with molecular clouds and produce high-energy radiation through proton-proton collisions \citep{Ackermann2013}. 
The HII region was formed by the ionization of hydrogen atoms. The generation of $\gamma$-rays in the HII region is a complex process involving a combination of ionization, non-thermal processes, nuclear reactions, and supernova explosions \citep{Niino2015}.
In addition, rare physical phenomena involving high-energy emissions, such as those from white dwarfs and red giants \citep{Cheung2022}, have been detected. 
White dwarfs or red giants usually do not produce high-energy radiation. However, in certain special cases, they may produce $\gamma$-ray emissions. For example, when a white dwarf or red giant undergoes explosive interaction with a companion star, a $\gamma$-ray burst may occur \citep{Cheung2022,Fryer1999,Barkov2010}. 

$\gamma$-rays are one of the high-energy electromagnetic waves, and their production mechanisms involve important high-energy physical processes, such as supernova explosions and black hole activity. By searching for new $\gamma$-ray sources, we can further understand the nature and mechanisms of these high-energy physical processes. 
Furthermore, $\gamma$-rays are a component of cosmic rays. By observing and analyzing them, we can study the origin and acceleration mechanisms of cosmic rays, which is essential for understanding the evolution and structure formation of the universe \citep{Blasi2013,Anchordoqui2003}. 
Given the significant value of $\gamma$-ray sources in astronomy, We utilize an efficient search approach to identify new $\gamma$-ray sources across the entire sky, leveraging advanced algorithms provided by the Fermipy \citep{Wood2017}.
 We applyed this method for the first time to analyze approximately 15.41 years of data from the Fermi-LAT all-sky survey, leading to the identification and cataloging of new $\gamma$-ray sources.
 In Section \ref{sec2}, we present the data analysis methods and results. Then, Section \ref{sec3} discusses and summarizes the results of the data analysis.

\section{Fermi-LAT data analysis} \label{sec2}
\subsection{\rm Data reduction } \label{sec2.1}

\begin{figure}
	\centering
	\includegraphics[width=16cm,height=9cm]{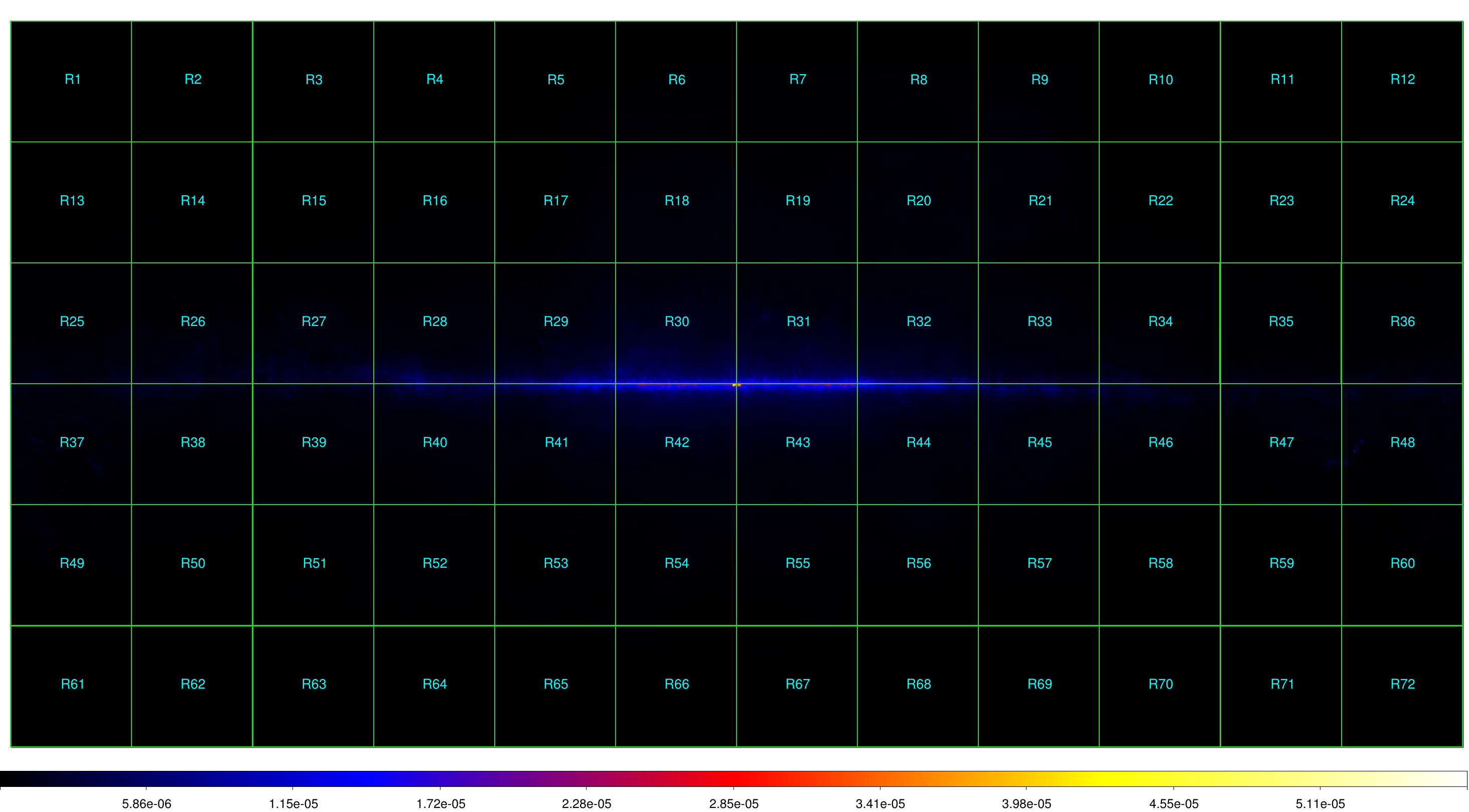}
	\caption{The map is the Galactic diffuse background emission, which had a size of 180$^{\circ}$ $\times$360$^{\circ}$, and its colored bar represents the radiation intensity in $\rm cm^{-2}\ s^{-1}\ sr^{-1}\ MeV^{-1}$. The whole region was  divided into 72 square regions, and the size of each region was 30$^{\circ}$ $\times$ 30$^{\circ}$. For details, see https://fermi.gsfc.nasa.gov/ssc/data/analysis/software/aux/4fgl/Galac-
	tic\_Diffuse\_Emission\_Model\_for\_the\_4FGL\_Catalog\_Analysis.pdf}
	\label{fig1}
\end{figure}

  To optimize the computational method used in this analysis, we employed the Galactic diffuse emission template as the background for the entire sky, covering 360$^{\circ}$ $\times$ 180$^{\circ}$. This area was divided into 72 regions of interest (ROIs), each measuring 30$^{\circ}$ $\times$ 30$^{\circ}$, as shown in Figure \ref{fig1}.
Our analysis used the latest version of the Fermipy, namely {\tt version 1.2.0}. The central coordinates of each region were individually chosen for  the binned likelihood analysis.   
We utilized the Pass 8 photon dataset of the P8R3{\_}V3 version for our analysis.
 The instrument response function used was  ``P8R3\_SOURCE\_V3'', and the commands ``evtype=3'' and ``evclass=128'' were employed to filter photon events. The number of energy bins per decade was set to binsperdec=10. The spatial bin size was set to ``binsz=0.1$^{\circ}$''. To avoid a larger point source spread function in the low-energy band, the photon energy range was selected to be from 500 MeV to 1 TeV. The selected time range was August 4, 2008 (MET 239557427) to December 29, 2023 (MET 725575162). The maximum zenith angle was set to  ``zmax=90$^{\circ}$'' to suppress cosmic-ray contamination from the Earth's limb.
Meanwhile, in the input model files required for the calculation, we included all sources from  the 4FGL-DR4\footnote{https://fermi.gsfc.nasa.gov/ssc/data/access/lat/14yr{\_}catalog/} within a 30$^{\circ}$ radius from the center of each region in Figure \ref{fig1}.   
We chose to free the prefactor and spectral index of all sources within 5$^{\circ}$ of each region of interest (ROI) shown in Figure \ref{fig1}; in addition, for two diffuse background radiation templates including the Galactic diffuse background emission ({\tt gll{\_}iem{\_}v07.fits}) and extragalactic isotropic diffuse background radiation ({\tt iso{\_}P8R3{\_}SOURCE{\_}V3{\_}v1.txt})\footnote{http://fermi.gsfc.nasa.gov/ssc/data/access/lat/BackgroundModels.html}, we freed their normalizations for fitting. We ensured that the fitting results in each region were optimal, with the fitting quality parameter {\tt fit{\_}quality=3}.

\subsection{\rm Search for new $\gamma$-ray sources}
In this study, we utilized the {\tt GTAnalysis} package from {\tt fermipy.gtanalysis} to load the results from Section \ref{sec2.1} using the {\tt load{\_}roi()} function. 
By referencing the significance level threshold from the 4FGL, we searched for new sources in each region using the {\tt find{\_}sources()} function, with a  significance threshold {\tt sqrt{\_}ts{\_}threshold}= 4.0, indicating that we selected results with significance levels of $\geq 4\sigma$.

Newly discovered sources were added to the input model and named in a form  similar to 4FGL-DR4, with the naming convention of the Fermi JHHMM.m+DDMM, based on their celestial coordinates\footnote{https://fermipy.readthedocs.io/en/latest/advanced/detection.html}.  
Subsequently, we utilized the {\tt localize()} function for positional improvement on each new source and determined their 1$\sigma$ and 2$\sigma$ position errors. The best-fit positions were adopted as the positions of the  new sources for all subsequent analyses. 
Considering that some faint $\gamma$-ray sources have TS values close to our selected threshold of 16, the TS values of the global fit are easily influenced by the  spectral parameters, resulting in TS values $<$ 16. Therefore, we freed the spectral parameters of the new sources with the best-fit positions and conducted a second binned likelihood fit using the {\tt fit()} function while ensuring that {\tt fit{\_}quality=3} for all fitting results and excluding sources with TS values $<$ 16.

\begin{figure}
	\centering
	\includegraphics[width=18cm,height=16cm]{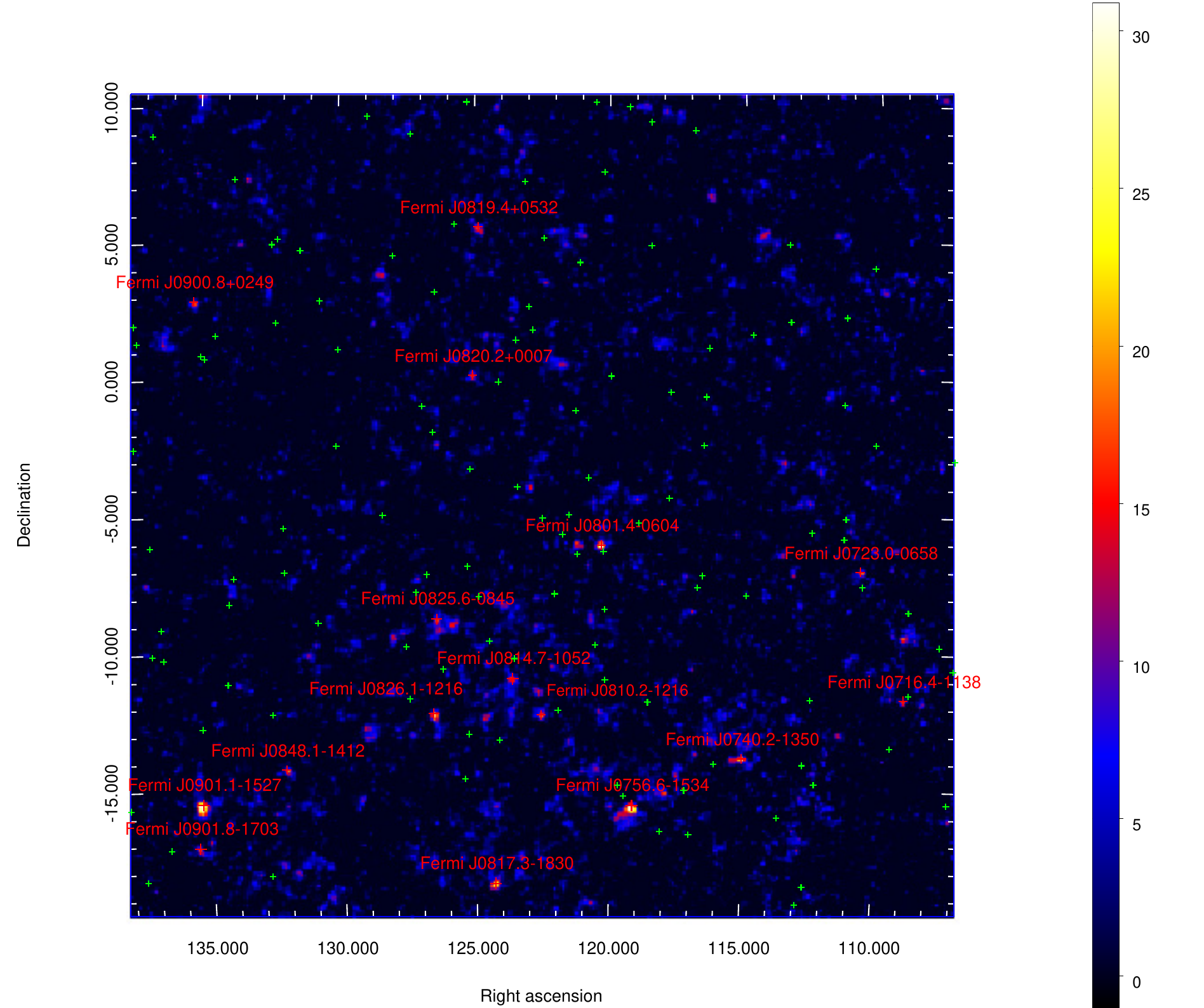}
	\caption{This map is the TS map of the R35 in Figure \ref{fig1}, which is 30$^{\circ}$ $\times$ 30$^{\circ}$ in size. The green crosses represent sources from 4FGL-DR4 and the red crosses represent the new sources discovered in this  study. The colored bar indicates TS values.}
	\label{fig2}
\end{figure}

Subsequently, the {\tt tsmap()} module was used to generate   TS maps with a size of 30$^{\circ}$ $\times$ 30$^{\circ}$ above 500 MeV. We loaded the positions of the new sources into the corresponding regions in Figure \ref{fig1} using the ds9 software\footnote{https://sites.google.com/cfa.harvard.edu/saoimageds9};   for example, the 35th region (R35) shown in Figure \ref{fig2}, from which shows that the locations of the new sources and those with TS values$>$16 strictly  coincide.

After collecting new sources from the 72 regions, it is essential to eliminate duplicate sources. The occurrence of duplicate sources arises because Fermipy's data analysis is based on the celestial coordinate system. During the data analysis process, the GTAnalysis method can only select a square region for likelihood analysis. This approach is consistent with the principle of creating a 3D binned counts map using the {\tt gtbin} command on the Fermi website.
Since the celestial coordinate system cannot be divided into uniform and consistently regular regions for analysis, the 72 square regions we selected in Fig 1 inevitably overlap when projected back onto the compressed celestial coordinate system. This overlap results in the occurrence of duplicate sources. To address this issue, we utilized the Excel functionality in WPS\footnote{https://www.wps.com/} to remove these duplicates. 
Finally, we compiled all the newly detected $\gamma$-ray sources into our catalog. Considering that this catalog is built upon the 4FGL-DR4, we named it 4FGL-Xiang.fits (4FGL-Xiang). Detailed information about this catalog can be found in Section \ref{sec5}, with explanations of relevant keywords available in the readme.txt file or Table \ref{Keynote}.

\subsection{\rm $\gamma$-ray spatial distribution analysis}

The {\tt extension()} function analyzed spatial extension for all new sources above 500 MeV. In this study, we calculated the spatial extension index TS$_{\rm ext}$ for each new source, where sources with TS$_{\rm ext}$ values $>$ 16 are considered to have significant extension features \citep{Acero2016}. The expression for TS$_{\rm ext}$ is $\rm TS_{ext} =2log( L_{ext}/L_{ps})$, as proposed by \citet{Lande2012}. 
We selected the two-dimensional Gaussian (Gauss) and uniform disk (Disk) \footnote{https://fermipy.readthedocs.io/en/latest/advanced/extension.html\#extension} as the spatial models. The radius (or $\sigma$) range of the tested models was set to 0.1$^{\circ}$-3.0$^{\circ}$, with increments of  0.01$^{\circ}$. We set the threshold of TS$_{\rm ext}$ to 16 and ensured that spatial models with TS$_{\rm ext}$ $>$ 16 were used as the best-fit spatial models for relevant sources in all subsequent analyses. This step was implemented by setting the {\tt sqrt{\_}ts{\_}threshold= 4.0} and {\tt update=True}. By testing two different spatial models, we collected a list of sources with TS$_{\rm ext}$ $>$ 16 in Table \ref{extTable3}.

\begin{table}
\begin{center}
\caption[]{Extended Source Catalog}\label{extTable3}

 \begin{tabular}{lllllll}
  \hline\noalign{\smallskip}
Name               & RA     & DEC    & Spatial\_model  & TS$_{\rm ext}$  &  ext & ext$_{\rm err}$ \\
  \hline\noalign{\smallskip}
Fermi J1016.5-5826e & 154.64 & -58.19 & Disk  & 22.89 & 1.32 & 0.04     \\
Fermi J0837.6-4759e & 129.75 & -48.03 & Disk  & 16.75 & 0.39 & 0.03     \\
Fermi J0830.7-4726e & 128.4  & -47.47 & Disk  & 17.64 & 1.34 & 0.05     \\
Fermi J0005.1+7334e & 0.46   & 73.37  & Disk  & 25.09 & 0.97 & 0.04     \\
Fermi J1821.1-1141e & 275.29 & -11.79 & Disk  & 25.74 & 1.70 & 0.14     \\
Fermi J1323.2-3911e & 200.92 & -39.19 & Disk  & 34.25 & 1.02 & 0.04     \\
Fermi J2247.5+5826e & 341.88 & 58.44  & Disk  & 58.84 & 1.32 & 0.01     \\
Fermi J1723.2-4035e & 260.81 & -40.59 & Disk  & 23.31 & 1.31 & 0.07     \\
Fermi J1030.9-5807e & 157.74 & -58.13 & Gauss & 19.01 & 0.86 & 0.14     \\
Fermi J1015.8-5825e & 153.66 & -58.5  & Gauss & 32.22 & 1.07 & 0.05     \\
Fermi J0830.6-4357e & 127.76 & -44.03 & Gauss & 50.12 & 2.49 & 0.25     \\
Fermi J0737.2-3212e & 114.41 & -32.21 & Gauss & 61.94 & 1.03 & 0.02     \\
Fermi J2140.1-4426e & 325.05 & -44.45 & Gauss & 33.98 & 2.95 & 0.10     \\
Fermi J0501.2+4434e & 75.4   & 44.64  & Gauss & 21.11 & 0.71 & 0.14     \\
Fermi J1504.1-5816e & 226.01 & -58.32 & Gauss & 23.51 & 0.58 & 0.09     \\
Fermi J1400.1-5825e & 209.85 & -58.2  & Gauss & 37.54 & 1.04 & 0.25     \\
Fermi J1405.4-5941e & 211.57 & -59.69 & Gauss & 43.19 & 0.97 & 0.15     \\
Fermi J1032.8-5756e & 158.29 & -57.93 & Gauss & 16.22 & 0.70 & 0.13     \\
Fermi J1908.5+0645e & 287.14 & 6.76   & Gauss & 29.46 & 0.63 & 0.10     \\
Fermi J1828.6-1542e & 277.15 & -15.7  & Gauss & 41.62 & 1.04 & 0.14     \\
Fermi J1714.6-4207e & 258.66 & -42.13 & Gauss & 16.51 & 1.08 & 0.04  \\
  \noalign{\smallskip}\hline
\end{tabular}
\end{center}
\end{table}

\subsection{\rm Spectral analysis} \label{sec:data-results}

The spectral curvature index TS$_{\rm cur}$, which is widely used to describe the curvature variability of a spectrum, is expressed as TS$_{\rm cur}$ =2(log L(curved spectrum) - log L(powerlaw)).  A source with a TS$_{\rm cur}$ $>$ 16 was considered significantly curved, as proposed by \cite{Nolan2012}.  
In our analysis, we utilized the {\tt curvature()} function from {\tt GTAnalysis} to calculate the TS$_{\rm cur}$ values, considering two commonly used spectral models: {\tt LogParabola} (LP) and {\tt PLSuperExpCutoff4} (PLE)\footnote{https://fermi.gsfc.nasa.gov/ssc/data/analysis/scitools/source{\_}models.html}.

	\begin{table*}[!h]
		\caption{Source with Spectral Curvature Variability \label{curTable4}}
		%\begin{center}
		\centering
			\begin{tabular}{lllll}
				\hline\noalign{\smallskip}
				Name            & RA      & DEC     & TS$_{\rm cur}$     & Spectral\_model \\
				\hline\noalign{\smallskip}	
Fermi J2218.7+6141  & 334.69 & 61.69  & 24.22 & LP  \\
Fermi J1813.3-1345 & 273.34 & -13.76 & 24.58 & LP  \\
Fermi J1821.1-1141e & 275.29 & -11.79 & 41.99 & LP  \\
Fermi J1705.2-4123 & 256.31 & -41.44 & 21.06 & LP  \\
Fermi J1717.8-4103  & 259.46 & -41.06 & 24.08 & LP  \\
Fermi J0948.7-3525 & 147.20 & -35.49 & 17.19 & LP  \\
Fermi J0631.3+1745 & 97.76  & 17.81  & 88.86 & LP  \\
Fermi J2116.7+4007  & 319.20 & 40.13  & 30.47 & LP  \\
Fermi J2032.7+3938  & 308.28 & 39.64  & 24.92 & LP  \\
Fermi J1833.4-1434 & 278.42 & -14.52 & 19.18 & LP  \\
Fermi J1714.6-4207e & 258.66 & -42.13 & 19.45 & LP  \\
Fermi J1403.6-6115  & 210.92 & -61.20 & 19.28 & LP  \\
Fermi J1405.4-5941e & 211.57 & -59.69 & 37.85 & LP  \\
Fermi J1046.3-5951  & 161.58 & -59.91 & 25.04 & LP  \\
Fermi J1051.4-5934  & 162.86 & -59.58 & 16.91 & LP  \\
Fermi J0830.7-4726e & 128.40 & -47.47 & 26.64 & LP  \\
Fermi J0535.3+2200  & 83.84  & 22.01  & 18.25 & LP  \\
Fermi J2151.1-4125  & 327.78 & -41.42 & 58.12 & LP  \\
Fermi J2146.2-4411  & 326.47 & -44.19 & 27.17 & LP  \\
Fermi J1801.6-0244  & 270.47 & -2.73  & 16.37 & PLE \\
Fermi J1814.1-1739  & 273.60 & -17.66 & 24.55 & PLE \\
Fermi J0636.3+1744  & 99.10  & 17.74  & 75.95 & PLE \\
Fermi J1527.4-6059 & 232.02 & -60.85 & 19.80 & PLE \\
				\noalign{\smallskip}\hline
			\end{tabular}
		%\end{center}
		%\label{Table1}
	\end{table*}

We calculated the TS$_{\rm cur}$ values of the two spectral models above 500 MeV for each new source. We determined whether these values were $>$ 16 and   compared the size of the TS$_{\rm cur}$ values of the two spectral
 models of each source. 
 We selected the spectral model with the maximum TS$_{\rm cur}$ value as the best-fit spectral model of the target source for all subsequent analyses and  listed the relative results in Table \ref{curTable4}.

\subsection{\rm Variability analysis} \label{sec:data-results}

In this study, we utilized the {\tt lightcurve()} function to generate the light curves of sources with TS values$>$25 above 1 GeV. We identified 361 sources and divided their light curves into ten time bins. We then calculated the variability index, TS$\rm _{var}$, for each object by referring to  \citet{Nolan2012}.  
For the light curves in the ten time bins, if their TS$\rm _{var}$ $>$  21.67\footnote{\textcolor{blue}{According to the description in the 4FGL \citep{Abdollahi2020}, TS$_{\rm var}$ follows a chi-square distribution. By utilizing the scipy.stats package \citep{Virtanen2020} to analyze the inverse cumulative distribution function, we determined that the threshold for TS$\rm _{var}$ at a 99\% confidence level is 21.67.}},  they are assumed to be a variable source with 99\% confidence. We identified 44 sources with variable characteristics, and the relevant information is presented in Table \ref{varTable}.

\section{Discussion and conclusion} \label{sec3}  

\subsection{\rm Description of Key Innovations} 

Utilizing the sophisticated algorithm incorporated in Fermipy, we introduce an efficient methodology for the comprehensive identification of $\gamma$-ray sources across the entire sky. 
Employing this methodology, we successfully identified 1,451 new $\gamma$-ray sources, each exceeding a significance level of 4$\sigma$. 
In contrast to the prior 4FGL-DR4 and earlier iterations, our results and methods are novel. Here we introduced the innovations through the following key aspects.

(a) More efficient search methodology.

Contrary to the analysis approaches previously recommended on the Fermi website, this paper presents a more efficient method specifically engineered for the identification of new sources. 
In the past, there have been two primary methods for identifying new sources. 
One method involves employing analysis techniques from  {\tt gtmodel} and {\tt farith} to examine the spatial distribution of residual photons within the  ROI. 
The second approach utilizes a TS map generated by the {\tt gttsmap} command to identify new sources.

The initial operational procedures for both methods are identical. Initially, users must select an ROI, followed by an analysis of all sources within the ROI using the likelihood method.
The first approach involves using {\tt gtbin} to create a photon counts map of the ROI, followed by the application of  {\tt gtmodel} to produce a model counts map based on the global fit.
Secondly, the FTOOL {\tt farith} is utilized to subtract the photon counts from the two maps, resulting in a photon residual map specifically designed for the ROI. 
Subsequently, users must conduct a visual inspection of the photon residual map of the ROI to identify new sources. For detailed procedural steps related to this method, please  refer to the Fermi website\footnote{https://fermi.gsfc.nasa.gov/ssc/data/analysis/scitools/binned{\_}likelihood{\_}tutorial.html}.

The first method enables visual assessment of the spatial distribution of residual photons; however, it cannot determine the significance level of $\gamma$-ray emissions at specific locations. This limitation is particularly  detrimental in identifying weak sources (4$\sigma$ to 5$\sigma$) amidst substantial background contamination.  
Consequently, to enhance the efficacy of identifying new sources, the second method has gained increasing acceptance.  Building on the binned likelihood analysis, this method employs {\tt gttsmap} to produce a TS map of the ROI.
By considering the quadratic relationship between the significance level and the TS value, we can effectively discern significant $\gamma$-ray emissions within the ROI. 
This methodology has been extensively applied in subsequent  analyses of Fermi-LAT data, facilitating the discovery of multiple significant $\gamma$-ray sources such as Arp 220 \citep{Peng2016}, SN 1006 \citep{Xing2016}, iPTF14hls \citep{Yuan2018}, and Kepler's SNR \citep{Xiang2021a}.

The two methods previously mentioned facilitate the discovery of new $\gamma$-ray sources; however, they possess a significant limitation: the requirement for manual identification via visual inspection. This manual process is particularly time-consuming, especially in regions with substantial background contamination.
If we identify the new sources depicted in Figure \ref{fig1} using the previously described methods, it necessitates  manual inspection of each location where the TS value exceeds 16 or there is an accumulation of a large number of photons, utilizing the ds9 software.

Given that visual identification may introduce unpredictable positional errors, it is crucial to utilize the {\tt gtfindsrc} command for the precise localization of each new source. These tedious procedures invariably require significant amounts of labor, computational resources, and time.
Fortunately, by leveraging advanced algorithms incorporated in Fermipy and our judiciously designed analytical procedures, we can significantly enhance the efficiency of identifying new sources.

The principal algorithm employed in this study is  the {\tt find{\_}sources()} function, which is similar to the second method described previously. Initially, it generates a TS map within an ROI. Then, it autonomously identifies positions where the TS value of each pixel exceeds the predefined threshold of 16, and subsequently places the model of point source at these locations for fitting.
This method supplants the manual inspection required by the two previous methods and furnishes critical information regarding the new $\gamma$-ray sources, including their positions, spectral parameters, fluxes, and TS values, among other details.
This method significantly optimizes the process of identifying new sources and enhances the efficiency of discovering new sources.
Moreover, to validate the optimality of the positions of new sources identified by the {\tt find{\_}sources()} function, we employed the {\tt localize()} function to test these positions. The results indicated no statistical difference between the two methods, confirming that the positions determined by {\tt find{\_}sources()} are highly accurate.

Furthermore, this study employed the Galactic diffuse background emission file as a reference to  partition the entire sky measuring 360$^{\circ}$ by 180$^{\circ}$. 
The Galactic diffuse background emission file is structured based on the Galactic coordinate system, allowing us to divide the entire sky into sections, each with a size constrained to 30$^{\circ}$ by 30$^{\circ}$.
Subsequently, we conducted a binned likelihood analysis simultaneously on the 72 regions depicted in Figure \ref{fig1}. By employing multi-threaded techniques, we further enhanced the efficiency of discovering new sources.

(b) More analytical data.

The dataset we utilized encompasses a period that exceeds that covered by the 4FGL-DR4. The 4FGL-DR4 dataset spans from 4 August 2008 to 2 August 2022, covering a duration of 14 years \citep{Ballet2024}.
In contrast, the dataset employed in this study extends from 4 August 2008 to 29 December 2023, covering 15.41 years. This represents an additional 1.41 years of data compared to the 4FGL-DR4 dataset.

(c) Discovery of 1379 new $\gamma$-ray sources.

\begin{figure}
	\centering
	\includegraphics[width=18cm,height=12cm]{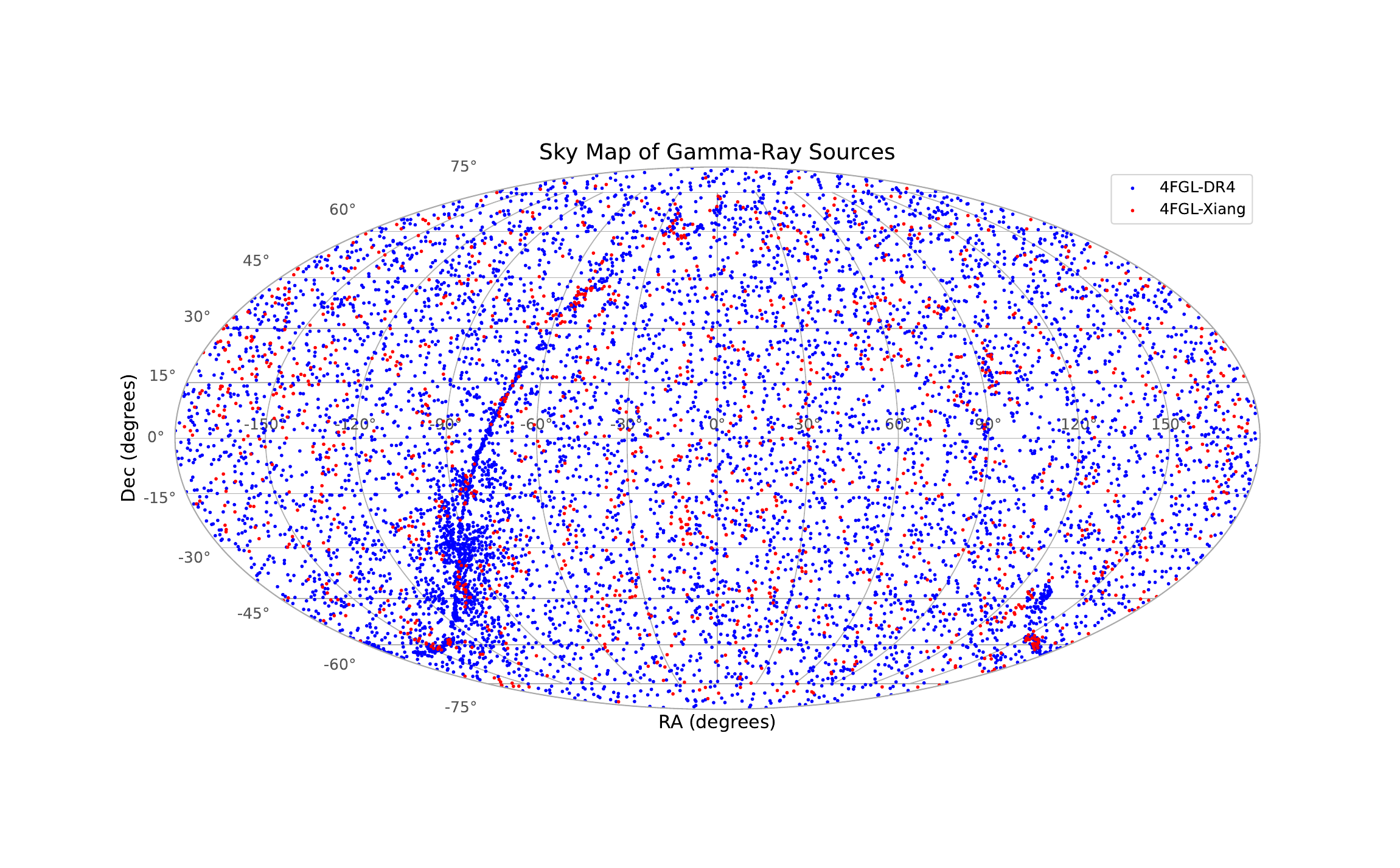}
	\caption{Sky Map of $\gamma$-Ray Sources.  Red dots represent data from 4FGL-Xiang and blue dots represent data from 4FGL-DR4.}
	\label{fig3}
\end{figure}

Firstly, it is necessary to note that the catalog utilized for generating the source model file in this study is the latest version, 4FGL-DR4\footnote{https://fermi.gsfc.nasa.gov/ssc/data/access/lat/14yr{\_}catalog/}.
To more intuitively display the spatial distribution of the new sources we detected and those in 4FGL-DR4, we plotted them together in spherical coordinates. As shown in Figure \ref{fig3}, it is evident that the new sources we detected are almost uniformly distributed across the entire sky.
Using the method mentioned in this paper, we identified 1379 new $\gamma$-ray sources. To avoid duplication with the sources listed in 4FGL-DR4, we conducted a targeted cross-identification between these new sources and those in 4FGL-DR4.

\begin{figure}
	\centering
	\includegraphics[width=16cm,height=16cm]{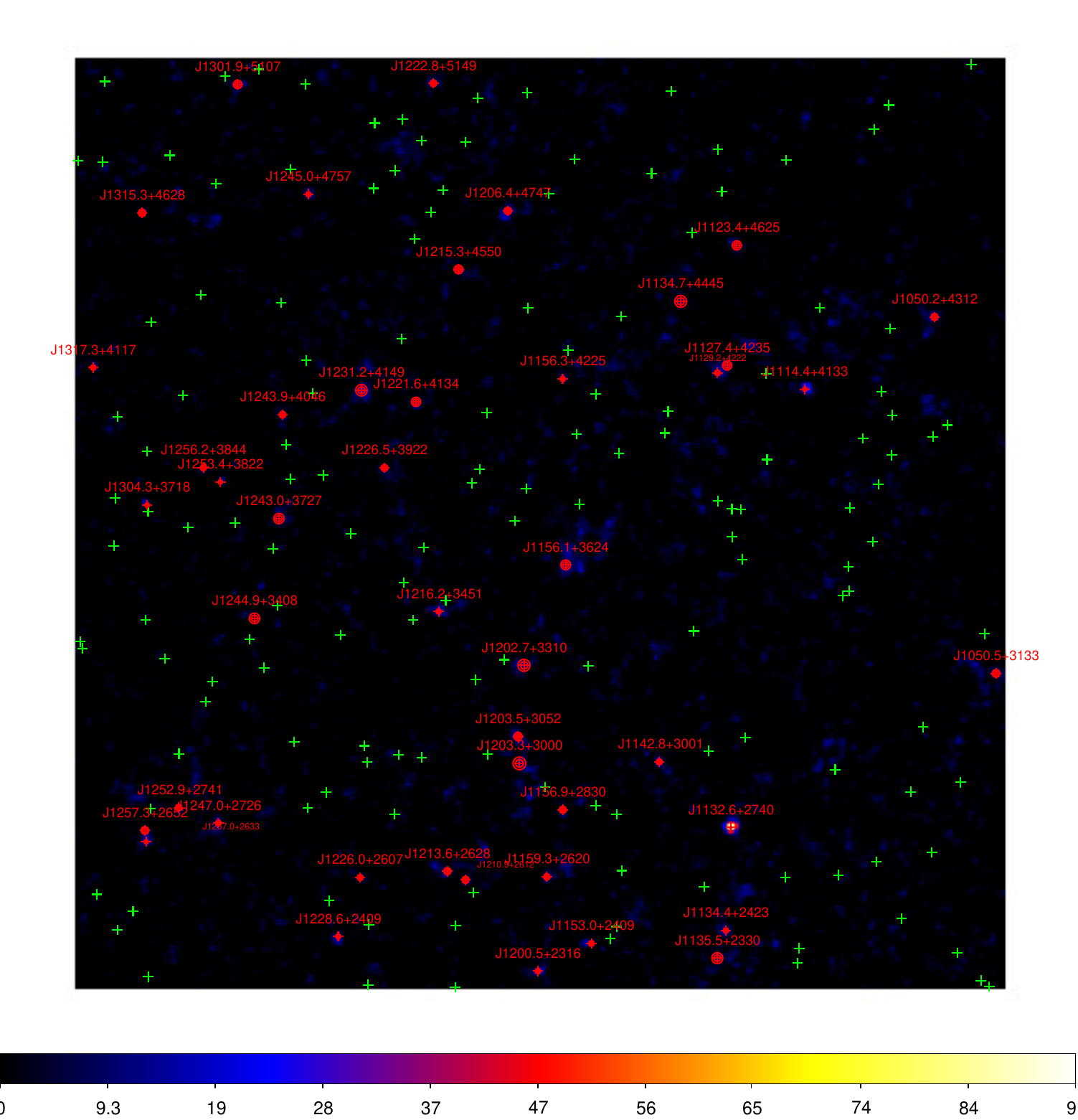}
	\caption{
This map is the TS map of R1 in Figure \ref{fig1}. The green and red crosses are described in Figure \ref{fig2}, where each new source is marked with two red circles representing the 1$\sigma$ and 2$\sigma$ positional errors, respectively.
}
	\label{fig4}
\end{figure}

To achieve this goal and help readers clearly identify the locations of these new sources relative to those in 4FGL-DR4, we constructed a new all-sky region file based on the latest released 4FGL-DR4 region file\footnote{https://fermi.gsfc.nasa.gov/ssc/data/access/lat/14yr\_catalog/}, named 4FGL-Xiang\_Region\_File.reg. This file includes the best-fit positions of all new sources, their 1$\sigma$ and 2$\sigma$ position errors, and all sources from 4FGL-DR4.
Additionally, we provided fits files of TS maps above 500 MeV, along with the corresponding  pre-generated eps files for 72 ROIs in Section \ref{sec5}. By using the following command (1), we can easily view the spatial positions of the new sources in each region. For example, the case for R1 is shown in Figure \ref{fig4},
\begin{equation}
$$\rm {\tt ds9\ run\_500MeV\_tsmap\_R*.fits\ -cmap\ b\ -region\ 4FGL-Xiang\_Region\_File.reg }$$ 
\label{eqx1}
\end{equation}
where * represents the ROI number. 
By examining the spatial positions of the new sources in the TS maps from 72 ROIs, we found no overlap between the sources in 4FGL-DR4 and the 2$\sigma$ error regions of the new sources. Additionally, the 2$\sigma$ error region of each new source generally contains some significant residual radiation. This strongly confirms the reliability of the $\gamma$-ray signals we detected.

By counting the significance levels of these new sources, we found that 4FGL-Xiang total includes  886 faint sources with a significance level between 4$\sigma$ and 5$\sigma$, and 508 significant $\gamma$-ray sources with a significance level greater than 5$\sigma$ above 500 MeV.

(d) Selecting energy range with a better angular resolution.

The current 4FGL-DR4 provides significance levels for $\gamma$-ray sources only within the energy range of 100 MeV to 1 TeV. Employing this single threshold for identifying new sources presents an evident drawback. Despite the broad energy coverage of this band, the large point-spread function at 100 MeV (approximately 5 degrees; see \cite{Abdollahi2020}) results in poor angular resolution, inevitably leading to significant background contamination.
Given the substantial accumulation of photons over a long period of 15.41 years across the entire sky, this exacerbates the background contamination caused by the large point-spread function in the lower energy band.
To mitigate significant background contamination, we chose a threshold of 500 MeV at the low energy end, which provides a better angular resolution of selected events than the previous analysis of 4FGL-DR4. 
Some analyses have demonstrated the benefits of employing a higher threshold at the lower energy range  when searching for new sources.
For example, \cite{Zhang2016} detected five $\gamma$-ray globular clusters above 400 MeV. \cite{Xin2019} verified high-energy emission from VER J2227+608 above 3 GeV. \cite{Xiang2021a,Xiang2021b}  confirmed the high-energy emission from Kepler's SNR and SNR G317.3-0.2 above 700 MeV and 2 GeV, respectively.

(e) Providing significance levels from more energy bands.

The 4FGL-DR4 does not provide information on significance levels for $\gamma$-ray sources across various energy ranges. This deficiency impedes targeted investigations of sources that emit significant high-energy radiation above 1 GeV or 10 GeV.
For example, when exploring SNRs emitting significant high-energy radiation above 10 GeV to analyze their multi-band non-thermal radiation characteristics, such as the work by \citet{Xin2019}, if we could promptly ascertain the significance level of VER J2227+608 above 10 GeV, it would enable us to determine the basic spectral properties of this object in advance. This would significantly enhance the efficiency of screening research samples.
Given the critical importance of significance levels across various energy bands in this field, this paper presents the significance levels for new sources within three distinct energy ranges, as detailed in 4FGL-Xiang: Sig$_{0.5}$ (0.5 GeV to 1 TeV), Sig$_{1}$ (1 GeV to 1 TeV), and Sig$_{10}$ (10 GeV to 1 TeV).

	\begin{table}[!h]
		\caption{Newly Discovered Variable Source \label{varTable}}
		%\begin{center}
		\centering
			\begin{tabular}{llll}
				\hline\noalign{\smallskip}
				Name               & RA     & DEC    & TS$\rm _{var}$ \\
				\hline\noalign{\smallskip}	
Fermi J1132.6+2740  & 173.14 & 27.71  & 172.08 \\
Fermi J1322.9+3215  & 200.73 & 32.25  & 24.31  \\
Fermi J1455.4+2134  & 223.82 & 21.57  & 24.12  \\
Fermi J1302.0+0534  & 195.52 & 5.57   & 49.21  \\
Fermi J1113.6+0457  & 168.42 & 4.95   & 749.49 \\
Fermi J0909.5+4251  & 137.40 & 42.87  & 210.06 \\
Fermi J0732.5+6156  & 113.15 & 61.95  & 22.01  \\
Fermi J1254.2-2001  & 193.61 & -19.99 & 82.98  \\
Fermi J2024.9+3959  & 306.29 & 39.99  & 25.02  \\
Fermi J1745.4+1721  & 266.41 & 17.32  & 24.25  \\
Fermi J1916.1+1041 & 288.99 & 10.69  & 28.40  \\
Fermi J1712.7-0500  & 258.26 & -5.02  & 33.67  \\
Fermi J1814.1-1739  & 273.60 & -17.66 & 49.94  \\
Fermi J1320.5-4522  & 200.17 & -45.39 & 28.35  \\
Fermi J1451.5-5240  & 222.90 & -52.67 & 23.69  \\
Fermi J1032.8-5756e & 158.32 & -57.95 & 32.91  \\
Fermi J0636.3+1744  & 99.10  & 17.74  & 49.58  \\
Fermi J0631.3+1745 & 97.76  & 17.81  & 86.54  \\
Fermi J0616.8+2222  & 94.21  & 22.38  & 26.00  \\
Fermi J1916.8+1141  & 289.21 & 11.70  & 30.84  \\
Fermi J1922.9+1413  & 290.72 & 14.20  & 56.97  \\
Fermi J1724.9-7257  & 261.43 & -72.96 & 25.77  \\
Fermi J1042.2-5937  & 160.56 & -59.67 & 37.70  \\
Fermi J1058.0-6044 & 164.51 & -60.75 & 37.29  \\
Fermi J0830.6-4357e & 127.76 & -44.03 & 24.20  \\
Fermi J0836.2-4438  & 129.09 & -44.68 & 36.58  \\
Fermi J0606.6+2029 & 91.68  & 20.50  & 31.03  \\
Fermi J0626.7+1734  & 96.78  & 17.69  & 35.07  \\
Fermi J0315.1+1010  & 48.85  & 10.17  & 24.40  \\
Fermi J0055.7+2106  & 14.06  & 21.15  & 45.57  \\
Fermi J0134.4+3049  & 23.62  & 30.83  & 31.04  \\
Fermi J2359.0+1028  & 359.77 & 10.47  & 31.94  \\
Fermi J0041.5+0833  & 10.42  & 8.60   & 58.70  \\
Fermi J2249.7+1304  & 342.43 & 13.13  & 32.11  \\
Fermi J2322.7-0155  & 350.65 & -1.92  & 80.88  \\
Fermi J2144.4+2047  & 326.12 & 20.78  & 36.41  \\
Fermi J2218.8-0328  & 334.71 & -3.52  & 36.95  \\
Fermi J2155.2-2536  & 328.82 & -25.61 & 43.17  \\
Fermi J2136.3-4440  & 324.09 & -44.67 & 30.92  \\
Fermi J0055.4-6122  & 13.86  & -61.43 & 53.25  \\
Fermi J0038.9-0121  & 9.76   & -1.37  & 30.70  \\
Fermi J2312.0-0503  & 347.96 & -5.07  & 937.95 \\
Fermi J2317.2-2556  & 349.45 & -25.97 & 40.79  \\
Fermi J0050.8-4229  & 12.76  & -42.50 & 57.33 \\
				\noalign{\smallskip}\hline
			\end{tabular}
		%\end{center}
		%\label{Table1}
	\end{table}

\subsection{\rm Verifying the robustness of the new method}\label{sec:compar_14_14}
Including an excessive number of degrees of freedom in parameter fitting can cause the binned likelihood fit to not converge, ultimately invalidating the binned likelihood analysis method. 
Therefore, to validate the robustness of the novel method, we conducted a comparative analysis of the fitting results for 556 $\gamma$-ray sources within a 5-degree radius centered at each IOR, as illustrated in Figure \ref{fig1}.

First of all, the period and model file utilized are strictly consistent with those published in 4FGL-DR4. Subsequently, we freed parameters within a 5-degree radius centered at each ROI illustrated in Figure \ref{fig1}.
We selected the energy range from 100 MeV to 1 TeV specifically to compare the significance levels with corresponding energy bands in 4FGL-DR4.
Furthermore, we compared the photon energy fluxes and their corresponding 1$\sigma$ errors above 100 MeV, derived using two distinct methods. The pertinent results are summarized in Table \ref{compar_14_14}. From this comparison, it was observed that the value of Ratio1 (defined as Sig$\rm _{14}$/Sig$\rm _{4FGL}$) is approximately 1, indicating that the values of Sig$\rm _{14}$ and Sig$\rm _{4FGL}$ are close. 
Additionally, no statistical differences were observed in the energy fluxes obtained from the two methods. The minor discrepancies in the analysis results between the methods could be attributed to inconsistencies in the hidden and free parameters. These findings substantiate the reliability of the newly designed method.

\begin{table*}[!h]
    \caption{\textbf{Analysis results of robustness verification} \label{compar_14_14}}
	\centering
	    \begin{tabular}{llllllllc}
			\hline\noalign{\smallskip}
				\hline\noalign{\smallskip}
		Name         & Sig$_{\text{14}}$ & Sig$_{\rm 4FGL}$ & Ratio1 & $\rm  Eflux_{\text{14}}$ & $\rm Unc\_Eflux_{\text{14}}$ &  Eflux & Unc\_Eflux & Offset\_{ROI} \\
		\hline\noalign{\smallskip}
		4FGL J1633.5+2806 & 5.13            & 4.95                  & 1.04                                  & 1.50E-12     & 3.12E-13          & 1.67E-12              & 3.48E-13                   & 4.10   \\
		4FGL J0925.7+3126 & 14.14           & 14.60                 & 0.97                                  & 3.51E-12     & 3.63E-13          & 3.25E-12              & 3.13E-13                   & 1.44   \\
		4FGL J0912.2+3004 & 4.21            & 4.39                  & 0.96                                  & 1.01E-12     & 2.90E-13          & 1.14E-12              & 3.02E-13                   & 2.83  \\
				... & ... & ... & ... & ... & ... & ... & ... & ...  \\
		\noalign{\smallskip}\hline
	   \end{tabular}
\begin{tablenotes} % 添加表格注释
            \item Note: Please refer to Table \ref{Keynote} for keyword explanations and units, and see Section \ref{sec5} for the full contents of the table.
\end{tablenotes}
\end{table*}

\subsection{\rm Exploring the $\gamma$-ray radiation characteristics of new sources}
Initially, we investigated the spatial extension properties of each new source and calculated their TS$_{\rm ext}$ values using the Gauss and Disk. We identified 21  sources with TS$_{\rm ext}$ $>$ 16. In comparison, the $\gamma$-ray spatial distributions of 13 sources was found to be consistent with the Gauss, and 8 sources were consistent with the Disk. The best-fit radius and $\sigma$  of the two models are presented in Table \ref{extTable3}. 
Subsequently, we investigated the spectral properties of the new sources using the LP and PLE spectral models. We calculated the TS$_{\rm cur}$ values for each model and identified 23 sources with significant curvature features with TS$_{\rm cur}$ $>$16 in their spectra. Among them, the spectra of 19 sources were found to be consistent with the LP, whereas those of four sources were consistent with PLE. The relevant results are summarized in Table \ref{curTable4}. 
Finally, analyzing the variability characteristics of bright sources is significant, for future studies on their flare mechanisms and quasiperiodic signals  \citep{Ulrich1997,Dermer2016,Madejski2016,Zhou2018,Ackermann2015,Benkhali2020}. We explored the light variation characteristics of bright sources with TS$>$25 above 1 GeV. 
The light curves of the 44 sources were variable with TS$\rm _{var}$ $>$21.67. The relevant results are given in Table \ref{varTable}. 
Using the results in Tables \ref{extTable3}, \ref{curTable4}, and  \ref{varTable}, we can conveniently obtain the $\gamma$-ray spatial distributions, spectral properties, and characteristics of the light variation of bright sources.

%他们的数量是170.　object of unknown Nature 数量是16.

4FGL-Xiang summarizes information on the $\gamma$-ray emissions of all new sources, with the header keywords described in Table \ref{Keynote}. 
With the continuous accumulation of Fermi-LAT data, it will be imperative to leverage it to track new sources in 4FGL-Xiang. This is particularly crucial for faint $\gamma$-ray sources with TS values ranging from 16 to 25 above 500 MeV, and it can aid in identifying more significant $\gamma$-ray sources and advancing our comprehension of critical issues, including the origin and evolution of the universe, particle acceleration mechanisms, high-energy radiation processes, and the genesis of cosmic rays.

\begin{table*}[!h]
    \caption{\textbf{Analysis results of robustness verification} \label{Keynote}}
	\centering
	    \begin{tabular}{lll}
			\hline\noalign{\smallskip}
				\hline\noalign{\smallskip}
		Column            & Unit    & Description \\
		\hline\noalign{\smallskip}
Name$^{a}$                    & ...     & New source name                                              \\
	RA                      & deg     & Right ascension                                          \\
	DEC                     & deg     & Declination                                              \\
	Spatial\_Model          & ...     & Best-fit spatial model                                   \\
	Spectral\_Model         & ...     & Best-fit spectral model                                  \\
	Index                   & ...     & Spectral index of powerlaw                                           \\
	Index\_err             & ...     & Spectral index error of powerlaw                              \\
	$F$               &$\rm cm^{-2}\ s^{-1}$ & Integral photon flux above 0.5 GeV                   \\
	$F_{\rm err}$          &$\rm cm^{-2}\ s^{-1}$ & Error of integral photon flux above 0.5 GeV          \\
	Np               & ...     & Predicted number of events in the model above 0.5 GeV                      \\
	Sig$_{0.5}$                 & $\sigma$     & The significance level above 0.5 GeV        \\
	Sig$_{1}$               & $\sigma$     & The significance level above 1 GeV               \\
	Sig$_{10}$              & $\sigma$     &The significance level above 10 GeV         \\
	TS$\rm _{cur}$               & ...     & Spectral curvature variability index   \\
	TS$\rm _{ext}$               & ...   & Spatial extension index                   \\
	Ext               & deg   & Spatial extension                                          \\
	Ext$_{\rm err}$          & deg     & Spatial extension error                                          \\	
	Pos$_{\rm r68}$                & arcsec  & Position error at 68\% confidence                          \\
	Pos$_{\rm r95}$                & arcsec  & Position error at 95\% confidence                          \\
	Type                    & ...     & Type of counterpart                                     \\
	Offset                 & arcsec  & Distance between new source and its counterpart                  \\
Sig$_{\rm 14}$ & $\sigma$  & The significance level above 100 MeV within 14 years   \\
Sig$_{\rm 4FGL}$& $\sigma$ & The significance level above 100 MeV within 14 years in 4FGL-DR4   \\
   	Ratio1 & ...     &  The ratio of Sig$_{\rm 14}$ to Sig$_{\rm 4FGL}$ \\
   
    Eflux &$\rm erg\ cm^{-2}\ s^{-1}$    &  Energy flux in 4FGL-DR4 \\
	Eflux$_{\rm 14}$& $\rm erg\ cm^{-2}\ s^{-1}$     & Energy flux above 100 MeV within 14 years \\
    
    Unc{\_}Eflux & $\rm erg\ cm^{-2}\ s^{-1}$    & 1$\sigma$ error on energy flux in 4FGL-DR4 \\
    Unc{\_}Eflux$\rm _{14}$ &  $\rm erg\ cm^{-2}\ s^{-1}$   & 1$\sigma$ error on energy flux above 100 MeV within 14 years \\
 
    Offset{\_}ROI& deg  & Angular offset from each ROI center \\
    N$\rm _{4FGL}$  & ...     & Predicted number of events  above 100 MeV in 4FGL-DR4 \\
		\noalign{\smallskip}\hline
	   \end{tabular}
\begin{tablenotes} % 添加表格注释
           \item Note:$^{a}$ {\tt e} represents the extended source.
\end{tablenotes}
\end{table*}

\subsection{\rm \textbf{Summary}}

  1.By leveraging the advanced algorithms offered by Fermipy, in conjunction with the Galactic diffuse background emission model and the imaging function of ds9, we devised an efficient method for identifying new $\gamma$-ray sources across the entire sky .

2.For the first time, we employed this method to identify $\gamma$-ray sources with significance level $>$ 4$\sigma$ across the entire sky. Through this approach, we discovered 1379 new $\gamma$-ray sources and conducted analyses of their locations, spectra, light curves, and spatial distributions. Meanwhile, the significance levels of three different energy bands are provided for future directional research.

3.We assessed the robustness of the new method within the parametric degrees of freedom permitted by the binned likelihood analysis.
Under the premise of maintaining consistency among the known parameters, we conducted a comparative analysis between the results of the new method and those from 4FGL-DR4. Our findings demonstrate that the results for the 556 sources involved in the analysis align closely with those from 4FGL-DR4, suggesting the reliability of the method.

\begin{acknowledgements}
We sincerely  appreciate the data and analysis software provided by the Fermi Science Support Center and also thank the support for this work from the Natural Science Foundation Youth Program of Sichuan Province (2023NSFSC1350), the Doctoral Initiation Fund of West China Normal University (22kE040), the Open Fund of Key Laboratory of Astroparticle Physics of Yunnan Province  (2022Zibian3), the Sichuan Youth Science and Technology Innovation Research Team (21CXTD0038), the National Natural Science Foundation of China (NSFC 12303048).
\end{acknowledgements}

\label{lastpage}


\begin{thebibliography}{99}
%% you can type \apj for ApJ, \aap for A&A, \apss for Ap&SS, etc. Please consult
%% the macro chjaa.cls. You can also find them in aasguide.tex (AASTeX for ApJ, AJ, PASP)
%% Please follow the format of ChJAA's reference list

\bibitem[Abdo et al.(2010)]{Abdo2010}Abdo, A. A., Ackermann, M., Ajello, M., et al., 2010, ApJS, 188, 405
\bibitem[Abdollahi et al.(2020)]{Abdollahi2020}Abdollahi, S., Acero, F., Ackermann, M., et al., 2020, ApJS, 247, 33

\bibitem[Acero et al.(2016)]{Acero2016}Acero,F., Ackermann,M., Ajello,M., et al., 2016,ApJS, 224,1

\bibitem[Ackermann et al.(2013)]{Ackermann2013}Ackermann, M. et al., 2013, Sci, 339, 807

\bibitem[Ackermann et al.(2015)]{Ackermann2015}Ackermann, M., Ajello, M., Albert, A., et al., 2015, ApJL,  813,2

\bibitem[Anchordoqui et al.(2003)]{Anchordoqui2003}Anchordoqui,L., Paul,T., Reucroft,S., et al., 2003,International Journal of Modern Physics A,  18,13

\bibitem[Ballet et al.(2024)]{Ballet2024}Ballet,J., Bruel,P., Burnett,T.H., \& Lott,B., 2024, arXiv:2307.12546

\bibitem[Barkov et al.(2010)]{Barkov2010}Barkov,M.V., Aharonian,F.A., Bosch-Ramon,V., 2010, ApJ, 724,2

\bibitem[Bell(1978)]{Bell1978}Bell, A. R., 1978, MNRAS, 182, 147

\bibitem[Benkhali et al.(2020)]{Benkhali2020}Benkhali, F. A., Hofmann, W., Rieger, F. M., et al., 2020, A\&A,  634: A120

\bibitem[Blasi(2013)]{Blasi2013}Blasi,P., 2013,The Astronomy and Astrophysics Review,  21, 1-73

\bibitem[Cheung et al.(2022)]{Cheung2022}Cheung,C.C., Johnson,T.J., Jean,P., et al., 2022, ApJ, 935,1

\bibitem[de Boer et al.(2017)]{deBoer2017}de Boer,W., Bosse,L., Gebauer,I., et al., 2017, PhRvD, 96,4

\bibitem[De Martino et al.(2013)]{DeMartino2013}De Martino, D., Belloni, T., Falanga, M., et al., 2013, A\&A,  550,A89

\bibitem[del Valle \& Romero(2014)]{delValle2014}del Valle, M.V., Romero,G.E., 2014,  A\&A, 563, A96

\bibitem[Dermer \& Giebels(2016)]{Dermer2016}Dermer, C. D., Giebels, B.,2016, Comptes Rendus Physique,  17,6

\bibitem[Fryer et al.(1999)]{Fryer1999}Fryer,C.L., Woosley,S.E., Herant,M., et al., 1999, ApJ, 520,2

\bibitem[Grondin et al.(2011)]{Grondin2011}Grondin,M.H., Funk,S., Lemoine-Goumard, M., et al., 2011, ApJ,  738,1

\bibitem[Goldstein et al.(2017)]{Goldstein2017}Goldstein,A., Veres,P., Burns,E., et al., 2017,  ApJL,  848,2

\bibitem[Gu\'{e}pin et al.(2018)]{Guepin2018}Gu\'{e}pin,C., Kotera,K., Barausse,E., et al., 2018, A\&A, , 616, A179

\bibitem[Gunn \& Ostriker(1969)]{Gunn1969}Gunn,J.E., Ostriker,J.P., 1969, Physical Review Letters,  22,14

\bibitem[Hartley(2017)]{Hartley2017}Hartley,W., 2017,   ApJL,  848, 2

\bibitem[Harvey et al.(2021)]{Harvey2021}Harvey,M., Rulten,C.B., Chadwick,P.M., 2021, arXiv preprint arXiv:2107.07215

\bibitem[Katsoulakos \& Rieger(2018)]{Katsoulakos2018}Katsoulakos,G., Rieger,F.M.,  2018, ApJ, 852,2

\bibitem[Lande et al.(2012)]{Lande2012}Lande, J., Ackermann, M., Allafort, A., et al., 2012, ApJ, 756, 5

\bibitem[Liu \& Yang(2022)]{Liu2022}Liu,B., Yang,R.,2022, A\&A, 659, A101

\bibitem[Madejski \& Sikora(2016)]{Madejski2016}Madejski, G., Sikora, M., 2016, Annual Review of Astronomy and Astrophysics,  54, 725-760

\bibitem[Narayan et al.(1992)]{Narayan1992}Narayan,R., Paczyński,B., Piran,T., 1992, arXiv preprint astro-ph/9204001
\bibitem[Niino et al.(2015)]{Niino2015}Niino,Y., Nagamine,K., Zhang,B., 2015,MNRAS, 449,3

\bibitem[Nolan et al.(2012)]{Nolan2012}Nolan, P. L., Abdo, A. A., Ackermann, M., et al., 2012, ApJS, 199, 31

\bibitem[Peng et al.(2016)]{Peng2016}Peng, F.-K., Wang, X.-Y., Liu, R.-Y., et al., 2016, ApJL, 821, L20

\bibitem[Peron et al.(2022)]{Peron2022}Peron,G., Casanova,S., Baghmanyan,V. et al., 2022, 7th Heidelberg International Symposium on High-Energy Gamma-Ray Astronomy, 7,394

\bibitem[Roberts et al.(2018)]{Roberts2018}Roberts,O.J., Fitzpatrick,G., Stanbro,M., et al., 2018, Journal of Geophysical Research: Space Physics,  123,5

\bibitem[Schlickeiser(1996)]{Schlickeiser1996}Schlickeiser,R., 1996, Space Science Reviews, 75, 299-315

\bibitem[Smith et al.(2023)]{Smith2023}Smith,D.A., Abdollahi,S., Ajello,M., et al., 2023,  ApJ,  958, 2

\bibitem[Strong (1977)]{Strong1977}Strong,A.W., 1977, Nature,  269,5627

\bibitem[Tchekhovskoy \& Giannios(2015)]{Tchekhovskoy2015}Tchekhovskoy,A., Giannios,D., 2015, MNRAS, 447,1

\bibitem[Ulrich(1997)]{Ulrich1997}Ulrich,M. H., Maraschi, L., Urry, C. M., 1997, Annual Review of Astronomy and Astrophysics,  35,1

\bibitem[Virtanen et al.(2020)]{Virtanen2020}
Virtanen, P., Gommers, R., Oliphant, T. E., et al., 2020, NatMe, 17, 261

\bibitem[Wood (2017)]{Wood2017}Wood, M., Caputo, R., Charles, E., et al., 2017, ICRC (Busan), 301, 824

\bibitem[Woosley \& Weaver(1995)]{Woosley1995}Woosley,S.E., Weaver,T.A., 1995, Lawrence Livermore National Lab.(LLNL), Livermore, CA (United States)

\bibitem[Woosley et al.(2002)]{Woosley2002}Woosley, S. E., Heger, A., Weaver,T. A., 2002, Reviews of modern physics,  74, 4

\bibitem[Woosley \& Bloom (2006)]{Woosley2006}Woosley, S. E., \& Bloom, J. S., 2006, ARA\&A, 44, 507

\bibitem[Xiang et al.(2021a)]{Xiang2021a}Xiang,Y.C., \& Jiang,Z.J., 2021, ApJ, 908, 22

\bibitem[Xiang et al.(2021b)]{Xiang2021b}Xiang,Y.C., Jiang,Z.J., \& Tang,Y.Y., 2021, ApJ, 911, 49

\bibitem[Xin et al.(2019)]{Xin2019}Xin, Y.-L., Zeng, H.-D., Liu, S.-M., Fan, Y.-Z., \& Wei, D.-M., 2019, ApJ, 885, 162

\bibitem[Xing et al.(2016)]{Xing2016}Xing, Y., Wang, Z., Zhang, X., \& Chen, Y., 2016, ApJ, 823, 44

\bibitem[Yuan et al.(2018)]{Yuan2018}Yuan, Q., Liao, N. H., Xin, Y. L., et al., 2018, ApJL, 854, L18

\bibitem[Zhang et al.(2008)]{Zhang2008}Zhang,L., Chen,S.B., Fang,J., 2008, ApJ,  676,2

\bibitem[Zhang et al.(2016)]{Zhang2016}Zhang, P. F., Xin, Y. L., Fu, L., et al., 2016, MNRAS, 459, 99

\bibitem[Zhou et al.(2018)]{Zhou2018}Zhou, J., Wang, Z., Chen, L., et al., 2018, Nature communications, 9,1


\end{thebibliography}
\end{document}